\documentstyle[12pt,epsf]{article}

\begin{document}

\def\fnum@figure{{\bf Figure. \thefigure}}

\long\def\@makecaption#1#2{
   \vskip 10pt 
   \setbox\@tempboxa\hbox{\small #1. #2}
   \ifdim \wd\@tempboxa >\hsize    
      \small #1. #2\par            
   \else                           
      \hbox to\hsize{\hfil\box\@tempboxa\hfil}  
   \fi}
    \setlength{\baselineskip}{2.6ex}

\centerline{\bf Pion-Nucleon Scattering in the Infinite Momentum Frame}
\vspace{8pt}

\centerline{Gerald. A. Miller}
\vspace{8pt}
\centerline{\em 
Physics Dep't, Univ. of Washington, Seattle, Washington 98195-1560}


\begin{abstract}
\setlength{\baselineskip}{2.6ex} A 
light front field theory treatment of a chiral lagrangian 
is applied to pion-nucleon and 
nucleon-nucleon scattering.  

\end{abstract}

\setlength{\baselineskip}{2.6ex}

\section*{Introduction}

Another title of this talk could be ``Light Front Treatment of Nuclei-Chiral 
Symmetry and Pion Nucleon scattering''.
A light front treatment is almost the  same as  the infinite 
momentum frame. 
   
The motivation for  this approach
 begins with a desire to understand  the EMC effect.
The observed 
 structure function 
depends on 
$x_{Bj}$, which in the parton model 
 is the ratio of the quark plus momentum to that of the target. If one regards
the nucleus as a collection of nucleons, $x_{Bj}
=p^+/k^+$, where  $k^+$ is the 
plus momentum of a bound nucleon.
A   
direct relationship between 
nuclear  theory and experiment occurs by
using a theory in which $k^+$ is one of the canonical variables.
Since $k^+$ is conjugate to a spatial variable  $x^-\equiv t-z$, 
it is natural to quantize the dynamical variables 
at the equal light cone time variable of $x^+\equiv t
+z$. This is  light front quantization. 
More generally, one expects to be able to profit from using
 light cone quantization in any situation which involves a 
 a large momentum.  

\section*{Redoing Nuclear Physics on the Light Front}

The use of a new quantization procedure requires that one redo all
of nuclear physics and check the most important features. 
The necessary steps include the following list.
 
\begin{enumerate}
\item Reproduce well-known results for nuclear matter in the mean field
approximation\cite{gam97a,gam97b}.
\item Provide a light front mean field treatment of
finite nuclei. P. Blunden 
(Manitoba) and I are undertaking this task now.
\item Provide a light front Bruckner theory of nuclear matter to include
the effects of correlations. R. Machleidt (Idaho) and I are working on this.
\item Obtain a light front chiral treatment of the NN force, 
for use in the Bruckner theory. 
\item Obtain a light front chiral treatment of $\pi$N  scattering.
\end{enumerate}
Here I present a low-order use of 
a chiral hadronic Lagrangian to 
obtain a light front treatment of pion-nucleon scattering. This  shows 
that the formalism works and indicates  it is possible to ultimately
obtain a fully relativistic chiral treatment of $\pi$N and NN 
 scattering.

\section*{Light Front Quantization}

Here is a primer aimed at providing the essentials of the
light front formalism.
One uses the energy momentum tensor $T^{\mu\nu}$ to construct the momentum
$P^\mu$ with the relation:
\begin{equation}
P^\mu={1\over 2}\int d^2x_\perp dx^- T^{+\mu}.
\end{equation}
The light front Hamiltonian is $P^-$. The nucleon fields $\psi$ are a 
four-component spinor, but there are only two independent degrees of freedom.
In the light front formalism it is easy to separate the 
dependent and independent variables using the projection operators
$\Lambda^\pm$ with 
$\Lambda\pm=\gamma^0(\gamma^0\pm\gamma^3)/2=\gamma^0\gamma^\pm.$
The independent field is $\psi_+=\Lambda_+\psi$, and the dependent field is
$\psi_-=\Lambda_-\psi$. The procedure is to use the field equations to obtain
$\psi_-$ as a function of $\psi_+$, and then express
$P^-$ in terms of independent fields.
 There is a similar treatment for vector mesons. The details of
the formalism can be found in  Refs.~\cite{gam97a,gam97b},  and especially the
references therein.

\subsection*{Lagrangian}

We use  a non-linear chiral
model in which the 
nuclear constituents are nucleons $\psi$ (or $\psi')$, pions ${\pi}$
 scalar mesons $\phi$ and
 vector mesons
$V^\mu$. 
The  pion and nucleon part of the Lagrangian ${\cal L}_{\pi N}$ is given by 
\begin{eqnarray}
{\cal L}_{\pi N}={1\over 4}f^2Tr (\partial_\mu U\partial^\mu
U^\dagger)+{1\over4}m_\pi^2f^2
Tr(U+U^\dagger-2)+\bar{\psi}^\prime\left(\gamma^\mu
({i\over 2}\stackrel{\leftrightarrow}{\partial}_\mu
-MU\right)\psi' \label{lag}
\end{eqnarray}
where the bare masses of the nucleon, scalar and vector mesons are given by 
$M, m_s,$  $m_v$, and  $V^{\mu\nu}=
\partial ^\mu V^\nu-\partial^\nu V^\mu$. The unitary  matrix $U$ can be
chosen as 
$U\equiv e^{i  \gamma_5 {\tau\cdot\pi}/f},\quad.$

The pion-nucleon  coupling here is chosen as that of  linear representations
of chiral symmetry used by  Gursey \cite{gursey}, with the 
 Lagrangian approximately 
invariant under a chiral transformation.
Another transformation can be used to convert this to
a Lagrangian of the non-linear
representation\cite{weinberg}. In this case
the early soft pion theorems are manifest in the Lagrangian, and the
linear pion-fermion coupling is of the pseudovector type.  However,
the use of light front theory, requires that one find an easy way to
solve the constraint equation that governs the fermion field. 
The constraint can be handled in a simple fashion by using
the linear representation\cite{gam97b}. 

Equation (\ref{lag})
may be thought of
as a low energy effective
theory for nuclei under normal conditions. 
Here we use its  light front Hamiltonian. 

The aim here is to understand low energy pion nucleon scattering. The effects
of pair suppression must emerge even though the linear pion-nucleon coupling
is given by the pseudoscalar $\gamma_5$ operator. 

\section*{Chiral Symmetry and Pion-Nucleon Scattering}

The first test for any chiral formalism is to reproduce the early soft pion
theorems.
Here we concentrate on low energy pion nucleon scattering
because of its relation to the nucleon-nucleon force. We work to second 
order in $1/f$ in this first application. In this case, $U$
takes the  form:
\begin{equation}
U=1+i\gamma_5 {{\tau\cdot\pi}\over f} -{\pi^2\over 2f^2}, \label{um1}
\end{equation}
which 
 is used to  construct the Hamiltonian $P^-$.
The second order scattering
graphs are of the three types  shown as time $x^+$ ordered 
diagrams in Fig.~1. The kinematics are such that
$\pi_i(q) N(k)\to \pi_f(q') N(k')$. 
 The direct and crossed graphs of Fig.~1a involve  
matrix elements of $\gamma_5$ 
between $u$ spinors, which vanish near threshold. The terms of Fig.~1b are
generated by  $\bar u \gamma_5 v$ terms. In any light front time-ordered
graph the 
plus-momentum is conserved, and the plus momentum of 
every  particle 
line  is greater than zero. This means that the first of Fig.1b 
vanishes identically, and the second vanishes for values of the initial 
pion plus momentum that are less than twice the nucleon mass.
 The net result is that only the instantaneous terms (which arise from
replacing $\psi_-$ by a function of $\psi_+$ )
and 
the $\pi^2$ 
term of  (shown in Fig. 1c) remain to be evaluated.
\begin{figure}[t]
\noindent
\epsfysize=3.0in
\hspace{1.2in}
\epsffile{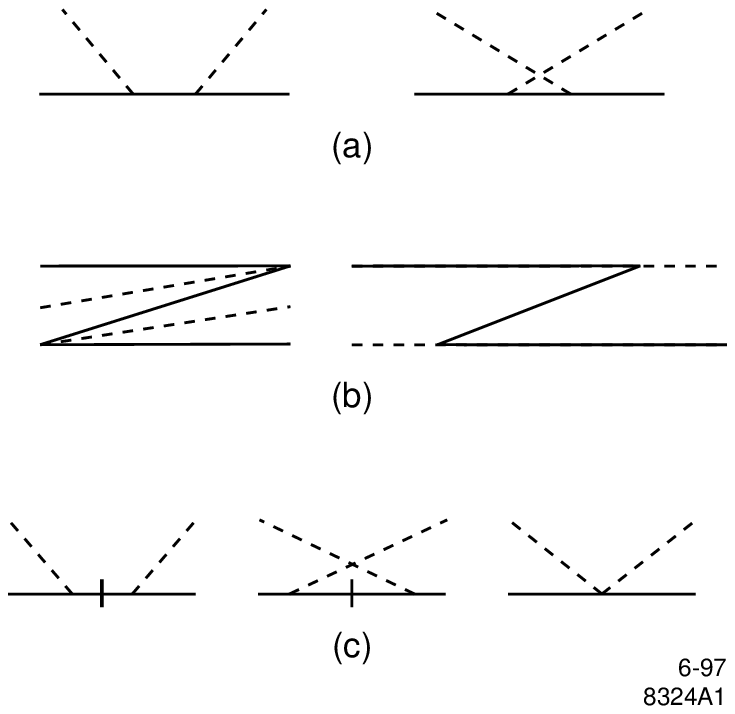}
\begin{center}
Fig.~1 $x^+$-ordered graphs for low energy pion-nucleon scattering.
\end{center}
\end{figure}

Proceeding more formally, the result is
\begin{equation}
{\cal M}=\tau_i\tau_f {M^2\over f^2} {\bar u(k')\gamma^+u(k)\over 2(k^++q^+)} +
\tau_f\tau_i {M^2\over f^2} {\bar u(k')\gamma^+u(k)\over 2(k^+-q^+)} 
-\delta_{if}{M\over f^2}\bar u(k')u(k)
\end{equation}
where the three terms here correspond to the three terms of Fig.~1c. The role
 of cancellations in the reduction of  the term proportional to $\delta_{if}$ 
is already apparent.
To understand the threshold physics take $k'^+=k^+=M$ and $q'^+=q^+=m_\pi$.
Then one finds 
\begin{equation}
{\cal M}= 
\delta_{if}{2m_\pi^2\over f^2} +2i\epsilon_{fin}\tau_n{m_\pi M\over f^2}
\end{equation}
to leading order in $m_\pi/M$. The weak nature 
of the $\delta_{if}$ term and the presence of the second Weinberg-Tomazowa term
is the hallmark of chiral symmetry.

\section*{Chiral Nucleonic Two  Pion Exchange Potential}
\begin{figure}[t] 
\noindent
\epsfysize=3.0in
\hspace{1.2in}
\epsffile{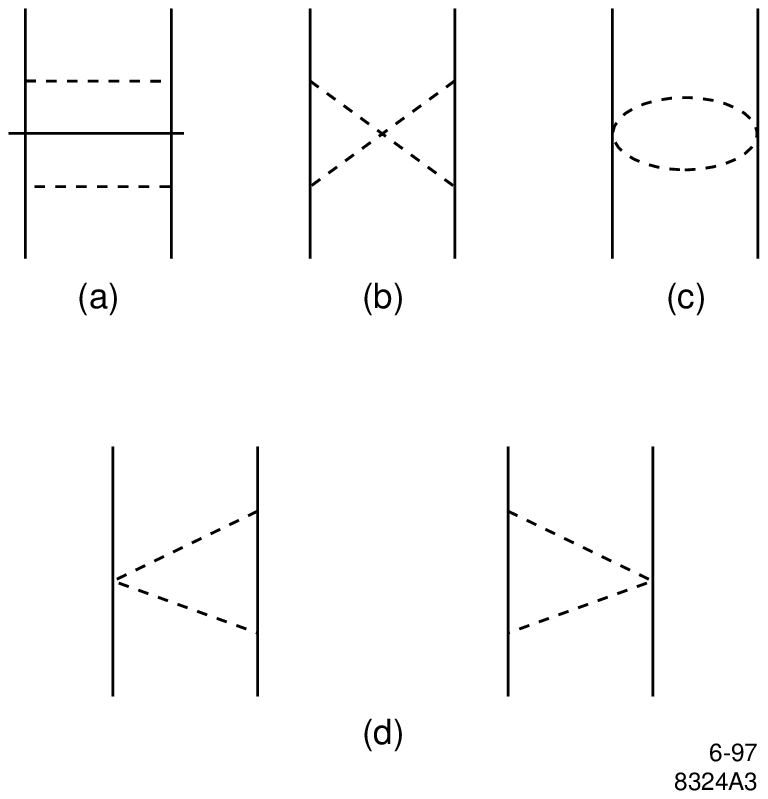}
\begin{center}
Fig.~2
Feynman graphs for the two-pion exchange potential.
\end{center}
\end{figure} 

We 
discuss the  two pion exchange contribution (of order $(M/f)^4$)
to the nucleon nucleon
potential.  The property that a sum of light cone
time-ordered diagrams  equals a single Feynman graph can be used to
simplify the calculation. The relevant Feynman graphs are displayed in
Fig.~2; the terms originating from the linear
$\gamma_5{\bf\tau}\cdot{\bf \pi}$ coupling (a,b), from the quadratic
$\pi^2-N$ coupling (c) and from a combination of the linear and
quadratic interactions (d) are indicated.  The line through the
two-nucleon intermediate state of Fig.~2a  indicates the subtraction of 
the contribution of the iterated  the one pion exchange
interaction. 

The sum of the terms of Fig.~2a and 2b is equal to the Partovi-Lomon
two pion exchange potential, as they  used the pseudoscalar pion-nucleon
interaction. This interaction certainly simplifies the calculation; in
particular the diagrams of Fig.~2a, b and d are convergent (whereas
they would be strongly divergent if pseudovector coupling were 
used).  One can use such a pseudoscalar coupling, and include the
effects of chiral symmetry, provided one also includes the effects of
the $\pi^2-N$ coupling shown in Fig.~2c, and the combined effects of
the linear and quadratic interactions, Fig.~2d.  The quadratic
interaction term cancels the large pair terms in pion-nucleon
scattering and should also play a significant role here in reducing
the size of the computed potential.  Thus we expect that the
Partovi-Lomon potential contains too large an attraction. 

Next turn to the procedure used in constructing the full Bonn
potential.  This potential is constructed by ignoring all of the
Z-graphs and including the the effects of the two-nucleon intermediate
states which arise from the crossed graph, Fig.~2b, as well as the
parts of Fig.~2a arising from time ordered terms in which two pions
exist at the same time. (For such contributions to the TPEP the linear
pseudoscalar and pseudovector interactions are are evaluated between
on shell positive energy nucleon spinors, and are therefore
equivalent.)  The resulting contribution to the TPEP is small, but is
comparable to that of the iterated OPEP.  The neglect of the Z graphs
goes a long way towards including the effects of chiral
symmetry. However, terms involving the Weinberg-Tomazowa interaction
at one or two vertices are ignored. The computation of the graphs of
Fig.~2 would include such effects implicitly as well as that of pair
suppression. Thus a detailed comparison would be useful. 
The small nature of the effects that we discuss now indicate that the 
dominance of the 
TPEP by effects of intermediate
$\Delta$'s will remain unchallenged.

\section*{Summary and Discussion}
The present paper argues that the light front quantization of a chiral
Lagrangian can handle 
pion-nucleon scattering. One can go beyond the present tree approximation and
obtain new relativistic chiral treatments of both pion-nucleon and nucleon
scattering. The resulting amplitudes can be used to obtain a new treatment
of nuclear physics which includes the effects of correlations and
relativity.

\bibliographystyle{unsrt}

\end{document}